\documentclass[draft,grl]{AGUTeX}

\usepackage{lineno}
\linenumbers*[1]

\usepackage[dvips]{graphicx}

\usepackage{amsmath,times,afterpage,color}
 \setkeys{Gin}{draft=false}
\authorrunninghead{CLAUDEPIERRE ET AL.}

\titlerunninghead{MAGNETOSPHERIC CAVITY MODES}

\authoraddr{S. G. Claudepierre and M. K. Hudson, Department of Physics and Astronomy, Dartmouth College, Hanover, NH 03755, USA. (Seth.G.Claudepierre@dartmouth.edu)}

\authoraddr{M. Wiltberger, NCAR, High Altitude Observatory, Boulder, CO 80301, USA.}

\authoraddr{S. R. Elkington, LASP, University of Colorado, Boulder, CO 80303, USA.}

\authoraddr{W. Lotko, Thayer School of Engineering, Dartmouth College, Hanover, NH 03755, USA.}

\begin{document}


\def\eq#1{Equation~(\ref{eq:#1})}
\def\fig#1{Figure~\ref{fig:#1}}
\def\tbl#1{Table~\ref{tbl:#1}}
\def\ch#1{Chapter~\ref{ch:#1}}
\def\sec#1{Section~\ref{sec:#1}}
\def\alfven{Alfv\'{e}n\ }

%
%

\title{Magnetospheric Cavity Modes Driven by Solar Wind Dynamic Pressure Fluctuations}
%

%
%

\authors{S. G. Claudepierre, \altaffilmark{1} 
M. Wiltberger, \altaffilmark{2}
S. R. Elkington, \altaffilmark{3} 
W. Lotko, \altaffilmark{4}
and M. K. Hudson \altaffilmark{1}}

\altaffiltext{1}{Department of Physics and Astronomy, Dartmouth College, Hanover, New Hampshire, USA}

\altaffiltext{2}{National Center for Atmospheric Research, High Altitude Observatory, Boulder, Colorado, USA}

\altaffiltext{3}{Laboratory for Atmospheric and Space Physics, University of Colorado, Boulder, Colorado, USA}

\altaffiltext{4}{Thayer School of Engineering, Dartmouth College, Hanover, New Hampshire, USA}



%
%
%

%
%


\begin{abstract}
We present results from Lyon-Fedder-Mobarry (LFM) global, three-dimensional magnetohydrodynamic (MHD) simulations of the solar wind-magnetosphere interaction.  We use these simulations to investigate the role that solar wind dynamic pressure fluctuations play in the generation of magnetospheric ultra-low frequency (ULF) pulsations.  The simulations presented in this study are driven with idealized solar wind input conditions.  In four of the simulations, we introduce monochromatic ULF fluctuations in the upstream solar wind dynamic pressure.  In the fifth simulation, we introduce a continuum of ULF frequencies in the upstream solar wind dynamic pressure fluctuations.  In this numerical experiment, the idealized nature of the solar wind driving conditions allows us to study the magnetospheric response to only a fluctuating upstream dynamic pressure, while holding all other solar wind driving parameters constant.  The simulation results suggest that ULF fluctuations in the solar wind dynamic pressure can drive magnetospheric ULF pulsations in the  electric and magnetic fields on the dayside.  Moreover, the simulation results suggest that when the driving frequency of the solar wind dynamic pressure fluctuations matches one of the natural frequencies of the magnetosphere, magnetospheric cavity modes can be energized.
\end{abstract}

%
%

%

\begin{article}

%

\section{Introduction \label{sec:intro}}

Several observational studies suggest that some dayside magnetospheric ultra-low frequency (ULF) pulsations may be directly driven by ULF fluctuations in the solar wind dynamic pressure.  For example, \citet{kepko:03a} examine six events where discrete ULF fluctuations are observed in the solar wind dynamic pressure.  The authors show a one-to-one correspondence between these solar wind dynamic pressure fluctuations and discrete spectral peaks in dayside GOES magnetic field data.  The authors argue that the dayside magnetospheric ULF pulsations are directly driven by the corresponding solar wind dynamic pressure fluctuations.  Other observational studies \citep{sibeck:89a,korotova:95a,matsuoka:95a,han:07a} also suggest that solar wind dynamic pressure fluctuations can directly drive dayside magnetic field ULF pulsations.  Very recent work \citep{viall:09a} concludes that approximately half of the variations observed in magnetospheric ULF waves are likely directly driven by solar wind dynamic pressure fluctuations.  In this study we investigate, through the use of global magnetohydrodynamic (MHD) simulations, the magnetospheric response to  ULF solar wind dynamic pressure (henceforth, $p_{dyn}$) fluctuations.  Here, `ULF' refers to frequencies in the 0.5 to 50 mHz  range (Pc3-Pc5 bands; \citet{jacobs:64a}), though we make no distinction between continuous and irregular magnetospheric pulsations.

\section{Methodology \label{sec:method}}

The details of the Lyon-Fedder-Mobarry (LFM) simulation code, the computational grid, and the numerical techniques used to solve the single-fluid ideal MHD equations can be found in \citet{lyon:04a}.  The solar wind input conditions form the outer boundary condition in the LFM simulation. For the inner boundary condition, the magnetospheric portion of the code couples to an empirical ionospheric model, which forms a two-way coupling between the simulation ionosphere and magnetosphere \citep{wiltberger:09a}.  The LFM simulation does not contain a plasmaspheric model and, thus, number densities in the simulation inner magnetosphere are lower than what is typically observed in the real magnetosphere. Also, as discussed in \citet{lyon:04a}, the LFM utilizes the Boris correction when solving the ideal MHD equations, where the speed of light is replaced by a smaller value to increase the allowable time step. The simulation code remains stable, however, when wave propagation speeds exceed the assumed speed of light, roughly 1,100 km/s in the LFM inner magnetosphere. We present results from five LFM simulations: four driven by monochromatic upstream $p_{dyn}$ fluctuations and one driven by a continuum of frequencies in the upstream $p_{dyn}$ fluctuations.

Solar wind dynamic pressure is not an explicit input  in the LFM simulation and we choose to introduce the dynamic pressure fluctuations via the upstream number density component, as opposed to the velocity component.  Solar wind observations typically show that $p_{dyn}$ variations are carried by the solar wind number density, and not the velocity \citep[e.g.][]{kepko:03a,han:07a}.  For the four monochromatic simulations, we impose a number density time series, $n(t)$, at the LFM upstream boundary at $x=\text{30}$ $R_E$ of the form:  $n(t)=n_{0}+\delta n \sin (\omega t)$.  The four monochromatic driving frequencies chosen for analysis in this study are 5, 10, 18, and 25 mHz and the background number density, $n_{0}$, is 5 particles/cm$^3$.  In the 5 and 10 mHz simulations, $\delta n= \text{1}$ (20$\%$ oscillation amplitude); in the 18 mHz simulation, $\delta n= \text{1.5}$ (30$\%$ oscillation amplitude); and in the 25 mHz simulation, $\delta n= \text{2}$ (40$\%$ oscillation amplitude).  The larger oscillation amplitudes for the input time series in the 18 mHz and 25 mHz runs are used to combat the effects of a numerical attenuation/filtering of higher frequency components in the LFM simulation.  For the fifth simulation, we impose a continuum of ULF frequencies in the input number density time series: $n(t)=n_{0}+0.05 \sum_{j} \sin (\omega_j t + \phi_j)$.  Here, we create an input spectrum with fluctuations in the 0 to 50 mHz band with a 0.1 mHz spacing between frequency components ($j$ ranges from 0 to 500) and we add a random phase, $\phi_j$, to each frequency component.  The value of 0.05 in the above equation is chosen so that the root-mean square ($RMS$) amplitude of the   continuum input number density time series is roughly equal to that of the monochromatic input number density time series (with 20$\%$ oscillation amplitudes).  In addition, in all five simulations, we introduce an appropriate out of phase oscillation in the input sound speed time series, so as to hold the thermal pressure constant in the upstream solar wind ($p_{th} \propto nC_{s}^{2}$). The background sound speed upon which the out of phase oscillation is imposed is 40 km/s.  The remaining idealized solar wind input parameters  are the same in all five simulations and held constant for the entire duration (4 hours) of the simulations:  {\bf B} = (0,0,-5) nT and {\bf v} = (-600,0,0) km/s.

The power spectral density ($PSD$) of the continuum simulation input $p_{dyn}$ time series is shown as the red trace in the inset panel in \fig{moneyplot}a.  Note the relatively uniform distribution of wave power over the 0 to 50 mHz frequency band.  The blue trace in the inset panel shows the $PSD$ of the $p_{dyn}$ time series taken at (20,0,0) $R_E$ in the solar wind (GSM coordinates are used throughout) in the continuum simulation.  Comparing the red and blue traces, we see that the spectral profile imposed at the upstream boundary (red trace) has been significantly altered by the time the fluctuations reach (20,0,0) $R_E$ (blue trace).  This filtering/attenuation of the higher frequency spectral components, to be discussed in a follow-up paper, is an expected artifact of the numerics in the LFM [{\it John Lyon, personal communication, 2008}].  Nonetheless, there is significant ULF wave power in the 0 to 20 mHz frequency band in the upstream $p_{dyn}$ driving, which is the spectral profile that drives the magnetosphere.  

The filtering/attenuation of the input time series in the continuum simulation results in upstream driving at (20,0,0) $R_E$ on the order of 13$\%$, reduced from the roughly 20$\%$ value imposed at the upstream boundary (in the $RMS$ sense described above).  As the inset panel in \fig{moneyplot}a suggests, the filtering/attenuation reduces the amplitude of the upstream $p_{dyn}$ driving at (20,0,0) $R_E$ to 24$\%$ in the 18 mHz simulation (input = 30$\%$) and 15$\%$ in the 25 mHz simulation (input = 40$\%$).  Finally, we note that upstream $p_{dyn}$ driving in the 13-24$\%$ range is reasonable when compared with the observational work discussed above and is at the lower end of what has been reported.

\section{Simulation Results \label{sec:simres}}

In all five simulations, the upstream $p_{dyn}$ fluctuations launch earthward propagating compressional MHD waves near the subsolar bow shock.  These waves propagate through the magnetosheath and then enter the magnetosphere near the subsolar magnetopause and propagate earthward through the dayside (not shown here).  We examine the magnetospheric response in the equatorial plane in terms of the compressional magnetic and electric field components, $B_z$ and $E_{\varphi}$.  Along the noon meridian, the magnetospheric response in terms of $B_z$ and $E_{\varphi}$ fluctuation amplitude is roughly an order of magnitude greater than in the other field components.  

\begin{center}
\begin{figure}
 \includegraphics[scale=0.3]{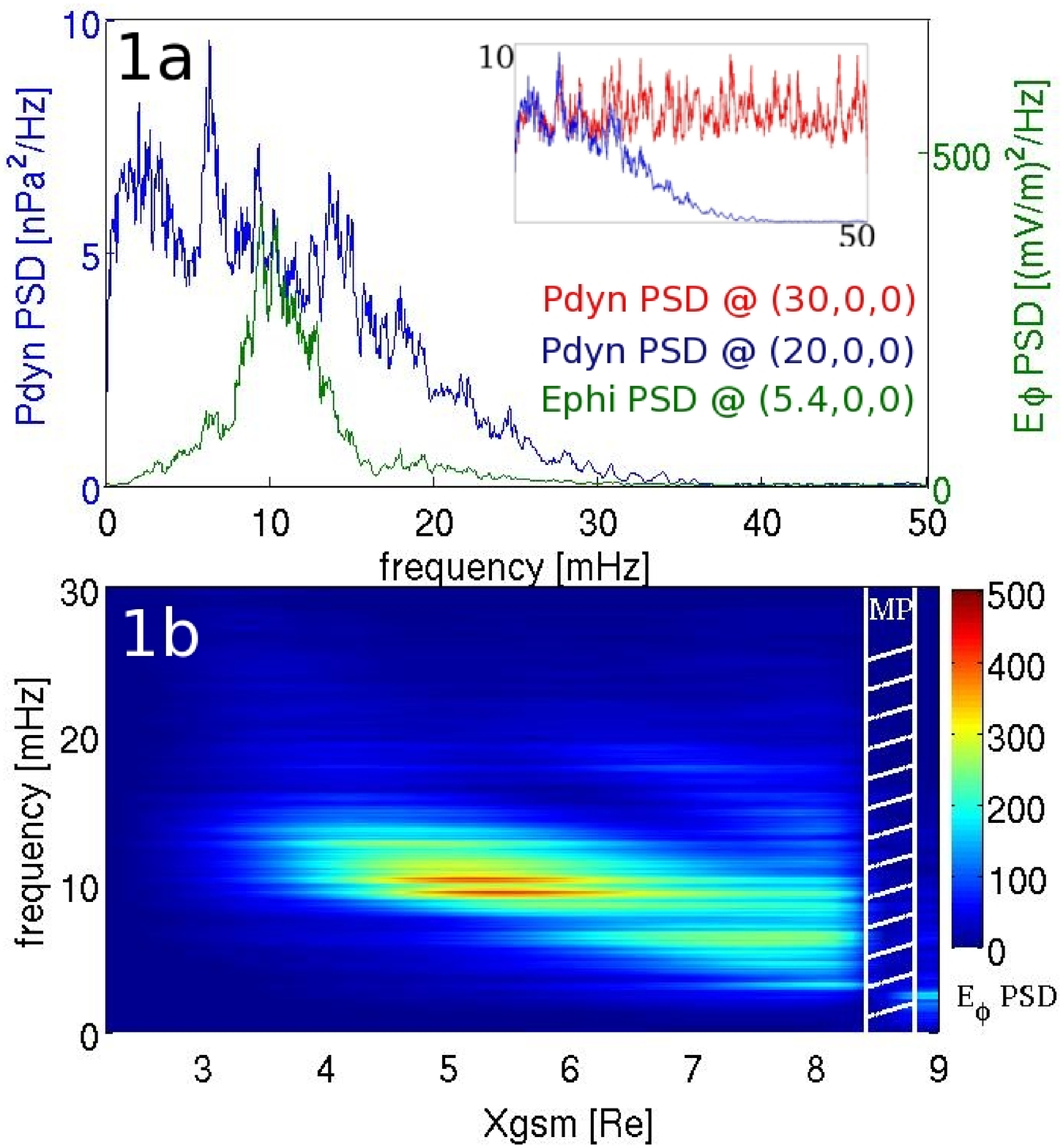}
 \caption{\label{fig:moneyplot} {\bf a)} Dynamic pressure PSD in the upstream solar wind (blue trace) and $E_{\varphi}$ PSD at 5.4 $R_E$ on the noon meridian (green trace), from the continuum simulation.  {\bf Inset panel:} $p_{dyn}$ PSD input at the LFM upstream boundary (red trace) and  $p_{dyn}$ PSD in the upstream solar wind (blue trace).  {\bf b)} $E_{\varphi}$ PSD plotted along the entire noon meridian in the continuum simulation.  The location of the magnetopause is indicated by the shaded region.}
\end{figure}
\end{center}

The green trace in \fig{moneyplot}a shows the magnetospheric response to the upstream $p_{dyn}$ fluctuations in the continuum simulation.  Here, we plot power spectral density of the $E_{\varphi}$ time series taken at (5.4, 0, 0) $R_E$ on the noon meridian.  Note the clear preferential frequency in the magnetospheric response centered near 10 mHz.  Comparing the fine structure in the spectral profile  of the $p_{dyn}$ fluctuations and the magnetospheric response  near 10 mHz shows a one-to-one correspondence between the two traces.  This suggests that the fluctuations in the magnetospheric $E_{\varphi}$ are driven by the $p_{dyn}$ fluctuations.  Moreover, the fact that the magnetospheric response is strongly peaked near 10 mHz suggests that the magnetosphere is responding resonantly to the upstream $p_{dyn}$ fluctuations, which contain a continuum of ULF frequencies.  Although the magnetospheric response falls off sharply away from 10 mHz, one could perhaps argue that the response near 10 mHz is due to local peaks in the upstream driving spectrum near 10 mHz.  The local peaks and valleys in the upstream driving spectrum are the result of the random phasing in the input time series and the discretization of the signal.  We have conducted analogous simulations to the continuum simulation presented here, with only the random phasing changed, which moves the location of the local peaks and valleys in the upstream $p_{dyn}$ driving spectrum.  These simulations also show a magnetospheric response that is strongly peaked near 10 mHz.  Thus, the magnetospheric response does not depend on the location of the local peaks and valleys in the upstream driving spectrum.

To obtain a more global picture of the magnetospheric response,  in \fig{moneyplot}b we plot the $E_{\varphi}$ PSD along the entire noon meridian in the continuum simulation.  Here, distance along the noon meridian is plotted on the horizontal axis from 2.2 $R_E$ (the inner boundary of the LFM simulation) to 9 $R_E$.  The subsolar magnetopause is located near 8.6 $R_E$ on the noon meridian, though the magnetopause moves roughly $\pm$0.25 $R_E$  about this location, due to the upstream $p_{dyn}$ oscillations.  This radial motion of the magnetopause is indicated by the shaded region in the figure.  Note that the green trace in \fig{moneyplot}a can be extracted from \fig{moneyplot}b by taking a vertical cut at 5.4 $R_E$.  The spectral profile along the entire noon meridian again shows a clear preferential frequency  near 10 mHz for the magnetospheric response.  Note that the frequency of the magnetospheric response does not change significantly with radial distance.  However, the amplitude of the  response near 10 mHz does depend on radial distance, with the maximum in wave power occurring between 5 and 6 $R_E$.  Finally, we note that there is an enhancement in the $E_{\varphi}$ PSD near 6 mHz, that peaks just earthward of the magnetopause, and decays rapidly in the earthward direction.  This is due to a local peak in the solar wind $p_{dyn}$ fluctuations near 6 mHz (\fig{moneyplot}a, blue trace)  and the fact that this local peak in the driving spectrum lies near a resonant frequency of the magnetosphere.

The results from the continuum simulation also suggest a secondary preferential frequency to the magnetospheric response, centered near 18 mHz.  However, the upstream driving in the continuum simulation near 18 mHz is weaker than the driving near 10 mHz, due to the filtering/attenuation described above.  Thus, the amplitude of the  secondary magnetospheric response is weaker than the primary response near 10 mHz, and is not entirely resolved in \fig{moneyplot}b due to the color scale used.   As we will see below, the amplitude of the secondary response near 18 mHz has two local maxima along the noon meridian, near 4 and 7 $R_E$, in contrast with one local maximum for the amplitude of the primary (10 mHz) response between 5 and 6 $R_E$.

In \fig{ripnoon}, we plot radial profiles of $E_{\varphi}$ (top row) and $B_{z}$ (bottom row) root-integrated power along the noon meridian for the five simulations in this study (columns).  Root-integrated power ($RIP$), plotted on the vertical axis in each of the 10 panels, is defined as: $RIP = (\int_{f_{a}}^{f_{b}} P(f) df)^{\frac{1}{2}}$, where $P(f)$ is the power spectral density of the time series under consideration and the integration is carried out over a given frequency band of interest, $[f_{a}, f_{b}]$.   In the four monochromatic simulations (\fig{ripnoon}, first four columns), the $RIP$ is integrated over the {\it driving band}, which we define as the 1 mHz frequency band centered on the driving frequency.  In the the continuum simulation (last column), two $RIP$ traces are shown, as there is no driving band in the continuum simulation.  The solid trace is integrated over the frequency band [7,12] mHz, to pick up the primary spectral peak near 10 mHz, while the dashed trace is integrated over the frequency band [15,20] mHz to pick up the secondary spectral peak near 18 mHz.  In each of the 10 panels, distance along the noon meridian is plotted on the horizontal axis and the location of the subsolar  magnetopause is indicated by the shaded regions near 8.5 $R_E$.

The five $B_z$ panels in the bottom row of \fig{ripnoon} show a strong amplitude maximum in $B_z$ oscillation amplitude near the magnetopause  that extends beyond the vertical scales used in the plots (the traces extend to a value on the order of 25 nT).  These strong oscillation amplitudes near the magnetopause are due to the radial motion of the magnetopause and the subsequent changing dayside magnetopause current.   As a side note, effects due to the LFM grid are clearly visible in the five $B_z$ panels in the bottom row of \fig{ripnoon}.  For example, in the 5 mHz simulation (bottom row, first panel) there is a `sawtooth' like structure in the radial profile between 5 and 7 $R_E$.  We do not attribute any physical significance to these features.  

\section{Discussion \label{sec:discuss}}

The simulation results presented above suggest a resonant response of the magnetosphere to solar wind dynamic pressure fluctuations, with a standing wave structure along the noon meridian.  The dependence of the magnetospheric response on the driving frequency can be explained by interpreting the simulation results as signatures of  magnetospheric cavity mode oscillations \citep[e.g.][]{kivelson:85a}.

\begin{center}
\begin{figure*}
 \includegraphics[scale=0.4]{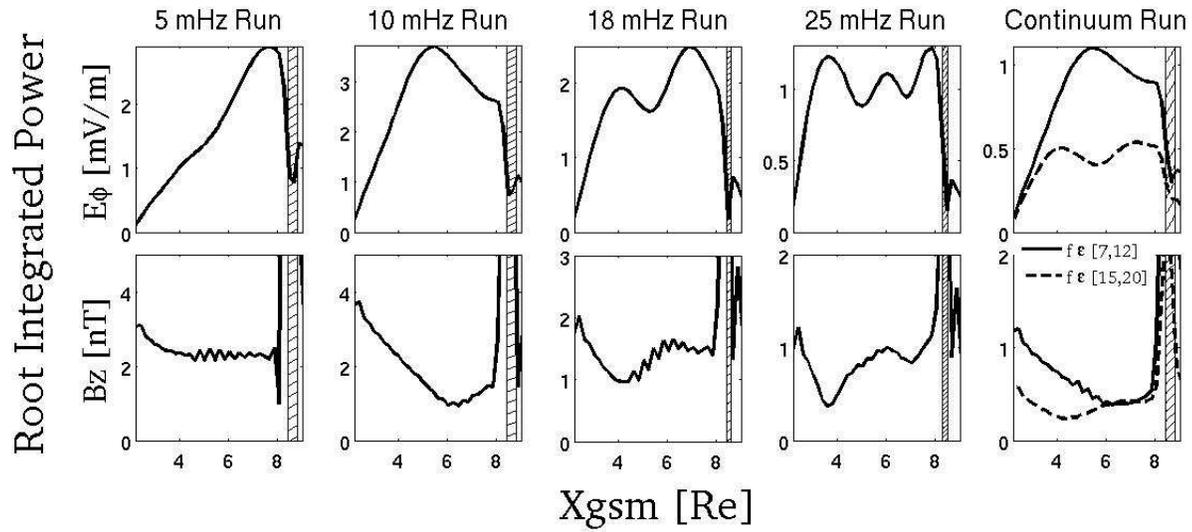}
 \caption{\label{fig:ripnoon} $E_{\varphi}$ (top row) and $B_z$ (bottom row) radial mode structure along the noon meridian for the five simulations (columns).  $RIP$ is integrated over the driving band in the monochromatic simulations (first four columns) and over [7,12] and [15,20] mHz in the continuum simulation (last column).  The location of the magnetopause is indicated by the shaded region.}
\end{figure*}
\end{center}

In the simplest interpretation, magnetospheric MHD cavity modes  can be thought of as standing waves in the electric and magnetic fields between a cavity inner and outer boundary.  We consider the magnetopause to be the cavity outer boundary and the LFM simulation inner boundary at 2.2 $R_E$ to be the cavity inner boundary.    For the moment, we consider perfect conductor boundary conditions at the simulation inner boundary and magnetopause ($E_{y}, \partial_{x} B_{z}\rightarrow 0$).  These boundary conditions impose half-wavelength standing waves in the radial direction between the simulation inner boundary and the magnetopause.  Returning to the noon meridian radial profiles in \fig{ripnoon}, we see that the simulation results support this standing wave   interpretation.  We argue that the $E_{\varphi}$ and $B_z$ radial profiles in the 10 mHz run (\fig{ripnoon}, second column) are the signatures of the $n=\text{1}$ cavity mode.  Near the simulation inner boundary and magnetopause, $E_{\varphi}$ has oscillation amplitude nodes and $B_z$ has oscillation amplitude antinodes.  Moreover, between the boundaries, $E_{\varphi}$ has  one oscillation amplitude antinode and $B_z$ has one oscillation amplitude node, near 6 $R_E$, all consistent with an $n=\text{1}$ standing wave along the noon meridian.  Note that the continuum simulation results suggest that the fundamental frequency of the magnetospheric cavity is near 10 mHz.  Thus, the upstream driving frequency in the 10 mHz monochromatic simulation is near the fundamental resonant frequency of the magnetospheric cavity and the $n=\text{1}$ radial eigenmode is excited.

In the 5 mHz simulation, we argue that a cavity mode is not excited, which is supported by the continuum simulation results.  The radial profile of  $E_{\varphi}$ along the noon meridian in the 5 mHz simulation (\fig{ripnoon}, top row, first column) suggests an evanescent decay of wave power, with $E_{\varphi}$ wave power peaking just earthward of the magnetopause and decaying rapidly in the earthward direction.  Monochromatic simulations with 1 mHz and 3 mHz driving, analogous to those presented here, show similar radial profiles in  $E_{\varphi}$ and $B_z$ oscillation amplitude along the noon meridian.  Thus, we argue that this is the characteristic behavior of dayside compressional magnetospheric disturbances under fluctuating solar wind $p_{dyn}$ driving when cavity modes are not excited, an evanescent decay of wave energy earthward of the magnetopause.  Finally, we note that the excitation of $n=\text{1}$ cavity mode in the 10 mHz simulation is also able to explain the stronger $E_{\varphi}$ response amplitude under 10 mHz monochromatic driving when compared with 5 mHz monochromatic driving.  The peak value of  $E_{\varphi}$ oscillation amplitude along the noon meridian is roughly 3.0 mV/m in the 5 mHz simulation, whereas it is roughly 3.7 mV/m in the 10 mHz simulation.  The only difference in the upstream driving in the two simulations is the driving frequency.  Thus, the magnetosphere responds resonantly to the $p_{dyn}$ fluctuations in the 10 mHz run and passively in the 5 mHz run.

We now consider the radial profiles of $E_{\varphi}$ and $B_z$ wave power in the 18 mHz and 25 mHz monochromatic simulations. We argue that in the 18 mHz simulation the $n=\text{2}$ cavity mode is excited.  Again, at the simulation inner boundary and magnetopause, $E_{\varphi}$ has oscillation amplitude nodes, whereas $B_z$ has amplitude antinodes.  Moreover,  near 4 and  7 $R_E$, $E_{\varphi}$ has two oscillation amplitude antinodes, whereas $B_z$ has two  nodes.  As discussed above, the continuum simulation results suggest that the frequency for an $n=\text{2}$ oscillation lies near 18 mHz, which is the driving frequency in the 18 mHz simulation.  Similarly, in the 25 mHz simulation, we argue that the $n=\text{3}$ cavity mode is excited.   In the  $E_{\varphi}$ profile, we see three clear oscillation amplitude antinodes near 4, 6 and 8 $R_E$.  Two of the three corresponding nodes in the $B_z$ profile are resolved near 4 and 7 $R_E$.  The $n=\text{3}$ interpretation  also requires a third $B_z$ node (likely between 4 and 7  $R_E$) that is not resolved in the simulation.  We note that the $B_z$  profile suggests that the $RIP$ value for the unresolved node should be less than 1 nT.  This would correspond to a peak-to-peak oscillation amplitude of roughly 2 nT or less, which is small when compared with background values on the order of 100's of nT.  The LFM grid resolution (roughly 0.25 $R_E$ here) coupled with the small oscillation amplitude may make it difficult to resolve three distinct $B_z$ nodes in an $\approx$3 $R_E$ range.  The $E_{\varphi}$ profile is consistent with the  $n=\text{3}$ cavity mode interpretation.  Finally, we note that the radial profiles from the continuum simulation, when integrated around 10 mHz (solid trace) and 18 mHz (dashed trace), look qualitatively similar to the profiles in the 10 mHz and 18 mHz monochromatic  simulations, respectively.  This suggests that the $n=\text{1}$ and $n=\text{2}$ radial eigenmodes are simultaneously excited in the continuum simulation.

The results from the continuum simulation suggest that the fundamental frequency of the magnetospheric cavity configuration is near 10 mHz.  To derive an alternate estimate, we consider the cavity frequency in a simple box geometry configuration \citep[e.g.][]{wright:94a}:
\begin{equation}
	f_{n} = \frac{V_{A}}{2a} n \quad \text{for } n = 1, 2, 3, \cdots
\label{eq:fcav}
\end{equation}
where $V_A$ is the \alfven speed in the box, $a$ is the box length in the $X$ direction, and $n$ is the quantization number.  Here, we envision the box coordinates, ($X$,$Y$,$Z$) as the radial, azimuthal and field aligned directions in the LFM.  The above equation for $f_{n}$ assumes perfect conductor boundary conditions in the $X$-direction ($E_{Y}, \partial_{X} B_{Z}\rightarrow 0$). To evaluate the fundamental frequency in the box configuration, we consider $n=1$ and only compare with LFM results from the 10 mHz and continuum simulations, as these are the only two simulations where the fundamental radial eigenmode is excited.  We evaluate the fundamental frequency, $f_{1}$, in the box with $a=\text{6.4 } R_E$, the distance from the simulation inner boundary to the magnetopause, along the noon meridian.  A value for the constant \alfven speed in the box, $V_A$, must also be chosen.  By tracking the compressional wave fronts  in the 10 mHz simulation, as they move earthward from the subsolar magnetopause along the noon meridian, we compute a phase speed, $V_{ph,x}$, of roughly 1,750 km/s.  With this estimate for $V_A$, we obtain $f_{1} \approx \text{22 mHz}$. For quarter-wavelength modes in the $X$ direction, the 2$a$ in \eq{fcav} is replaced by 4$a$ and the fundamental cavity frequency is $f_{1} \approx \text{11 mHz}$, close to the result suggested by the continuum simulation. The fact that the quarter-wavelength fundamental cavity frequency is closer to 10 mHz than the half-wavelength estimate and the fact that the electric field oscillation amplitude does not go entirely to zero at the magnetopause both suggest that quarter-wavelength modes may be a more appropriate boundary condition at the magnetopause.

As discussed in \sec{method}, the speed of light in the LFM is set to an artificially low value, which limits the \alfven wave propagation speed.  Above, we computed a phase speed of roughly 1,750 km/s for the $p_{dyn}$-driven waves in the 10 mHz simulation, which exceeds the speed of light in the simulation.  Thus, the wave propagation characteristics of the $p_{dyn}$-driven waves are effected by the Boris correction.   The Boris correction compensates for neglect of the displacement current in the ideal MHD equations, which reduces the phase speed when $V_A \sim c$.

Finally, we emphasize that the results presented in this study do not necessarily imply that the fundamental cavity frequency of the real magnetosphere is near 10 mHz.  A key factor controlling the fundamental frequency of the magnetospheric cavity is the \alfven speed profile.  The LFM simulations presented in this study do not have a plasmaspheric model and, thus, have number densities in the dayside equatorial plane that are much lower than in the real magnetosphere.  For example, a typical value for the LFM number density near (5,0,0) $R_E$ is 0.1 particles/cm$^{3}$.  A more realistic LFM number density profile in the equatorial plane, under development, would  significantly lower the fundamental cavity frequency of the LFM magnetosphere (e.g. \eq{fcav}).  For this reason, we do not compare the LFM simulation results with the observations of magnetospheric ULF waves driven by $p_{dyn}$ fluctuations discussed in \sec{intro}.  The observational work \citep[e.g.][]{kepko:03a} typically looks at frequencies less than 5 mHz,  while we have shown that the lowest cavity mode frequency that the LFM supports, for these upstream parameters, is approximately 10 mHz.

%
%
%
%
%
%

%
%
%
%

\begin{acknowledgments}
The authors are grateful for thoughtful discussions with R. E. Denton and J. G. Lyon.
\end{acknowledgments}


\begin{thebibliography}{11}
\providecommand{\natexlab}[1]{#1}
\expandafter\ifx\csname urlstyle\endcsname\relax
  \providecommand{\doi}[1]{doi:\discretionary{}{}{}#1}\else
  \providecommand{\doi}{doi:\discretionary{}{}{}\begingroup
  \urlstyle{rm}\Url}\fi

\bibitem[{\textit{Han et~al.}(2007)}]{han:07a}
Han, D.~S., et~al. (2007), Coupling of perturbations in the solar wind density
  to global {Pi}3 pulsations: {A} case study, \textit{J.\ Geophys.\ Res.},
  \textit{112}(A05217).

\bibitem[{\textit{Jacobs et~al.}(1964)\textit{Jacobs, Kato, Matsushita, and
  Troitskaya}}]{jacobs:64a}
Jacobs, J.~A., Y.~Kato, S.~Matsushita, and V.~A. Troitskaya (1964),
  Classification of geomagnetic micropulsations, \textit{J.\ Geophys.\ Res.},
  \textit{69}, 180.

\bibitem[{\textit{Kepko and Spence}(2003)}]{kepko:03a}
Kepko, L., and H.~E. Spence (2003), Observations of discrete, global
  magnetospheric oscillations directly driven by solar wind density variations,
  \textit{J.\ Geophys.\ Res.}, \textit{108}(A6), 1257,
  \doi{10.1029/2002JA009676}.

\bibitem[{\textit{{Kivelson} and {Southwood}}(1985)}]{kivelson:85a}
{Kivelson}, M.~G., and D.~J. {Southwood} (1985), {Resonant ULF waves - A new
  interpretation}, \textit{Geophys.\ Res.\ Lett.}, \textit{12}, 49--52,
  \doi{10.1029/GL012i001p00049}.

\bibitem[{\textit{Korotova and Sibeck}(1995)}]{korotova:95a}
Korotova, G.~I., and D.~G. Sibeck (1995), A case study of transient event
  motion in the magnetosphere and in the ionosphere, \textit{J.\ Geophys.\
  Res.}, \textit{100}(A1), 35--46.

\bibitem[{\textit{Lyon et~al.}(2004)\textit{Lyon, Fedder, and
  Mobarry}}]{lyon:04a}
Lyon, J.~G., J.~A. Fedder, and C.~M. Mobarry (2004), The
  {L}yon--{F}edder--{M}obarry {(LFM)} global {MHD} magnetospheric simulation
  code, \textit{J.\ Atmos.\ Solar-Terr.\ Phys.}, \textit{66}(15), 1333,
  \doi{10.1016/j.jastp.2004.03.020}.

\bibitem[{\textit{Matsuoka et~al.}(1995)\textit{Matsuoka, Takahashi, Yumoto,
  Anderson, and Sibeck}}]{matsuoka:95a}
Matsuoka, H., K.~Takahashi, K.~Yumoto, B.~J. Anderson, and D.~G. Sibeck (1995),
  Observation and modeling of compressional {Pi} 3 magnetic pulsations,
  \textit{J.\ Geophys.\ Res.}, \textit{100}(A7), 12,103--12,115.

\bibitem[{\textit{{Sibeck} et~al.}(1989)\textit{{Sibeck}, {Baumjohann},
  {Elphic}, {Fairfield}, and {Fennell}}}]{sibeck:89a}
{Sibeck}, D.~G., W.~{Baumjohann}, R.~C. {Elphic}, D.~H. {Fairfield}, and J.~F.
  {Fennell} (1989), {The magnetospheric response to 8-minute period
  strong-amplitude upstream pressure variations}, \textit{J.\ Geophys.\ Res.},
  \textit{94}, 2505--2519, \doi{10.1029/JA094iA03p02505}.

\bibitem[{\textit{{Viall} et~al.}(2009)\textit{{Viall}, {Kepko}, and
  {Spence}}}]{viall:09a}
{Viall}, N.~M., L.~{Kepko}, and H.~E. {Spence} (2009), {Relative occurrence
  rates and connection of discrete frequency oscillations in the solar wind
  density and dayside magnetosphere}, \textit{J.\ Geophys.\ Res.},
  \textit{114}(A13), 1201--+, \doi{10.1029/2008JA013334}.

\bibitem[{\textit{Wiltberger et~al.}(2009)\textit{Wiltberger, Weigel, Lotko,
  and Fedder}}]{wiltberger:09a}
Wiltberger, M., R.~S. Weigel, W.~Lotko, and J.~A. Fedder (2009), {Modeling
  seasonal variations of auroral particle precipitation in a global-scale
  magnetosphere-ionosphere simulation}, \textit{J.\ Geophys.\ Res.},
  \textit{114}(A13), 1204--+, \doi{10.1029/2008JA013108}.

\bibitem[{\textit{{Wright}}(1994)}]{wright:94a}
{Wright}, A.~N. (1994), {Dispersion and wave coupling in inhomogeneous MHD
  waveguides}, \textit{J.\ Geophys.\ Res.}, \textit{99}, 159--167,
  \doi{10.1029/93JA02206}.

\end{thebibliography}

%
%
%
%
%
%
%
%
%
%


%
%

\end{article}




%
%
%
%
%
%


\end{document}